\documentclass[aps,prd,superscriptaddress,twocolumn,floatfix,showkeys]{revtex4}

\usepackage{amsfonts}

\usepackage{tikz}
\usepackage{tikz-3dplot}

\usepackage{amsmath}
\usepackage{amssymb}
\usepackage{amsfonts}
\usepackage{bbold}
\usepackage{bm}
\usepackage{cancel}
\usepackage{times,float}

\usepackage{float}
\usepackage{graphicx}
\graphicspath{{FIGURES/}}
\usepackage[dvipsnames,svgnames]{xcolor}
\usepackage{hyperref}
\hypersetup{colorlinks=true, linkcolor=Blue, citecolor=Blue,urlcolor=Blue}
\usepackage{multirow}
\usepackage{ulem}
\usepackage{color,epsfig}
\usepackage{bm}
\usepackage{slashed}
\usepackage{hyperref}
\usepackage{feynmf}
\usepackage{array}
\usepackage{physics}
\usepackage{subcaption}
\allowdisplaybreaks
\usepackage{comment}
\usepackage[none]{hyphenat}

\begin{document}
\tdplotsetmaincoords{70}{110}

\sloppy 

\title{
Spontaneous persistent currents and time-reversal symmetry breaking in thick-walled Weyl semimetal cylinders}

\author{J. C. Pérez-Pedraza}
\email{julio.perez@correo.nucleares.unam.mx}
\affiliation{Instituto de Ciencias Nucleares, Universidad Nacional Aut\'{o}noma de M\'{e}xico, 04510 Ciudad de M\'{e}xico, M\'{e}xico}

\author{Juan A. Cañas}
\email{juan.canas@correo.nucleares.unam.mx}
\affiliation{Instituto de Ciencias Nucleares, Universidad Nacional Aut\'{o}noma de M\'{e}xico, 04510 Ciudad de M\'{e}xico, M\'{e}xico}

\author{Daniel A. Bonilla}
\email{daniel.bonillam@correo.nucleares.unam.mx}
\affiliation{Instituto de Ciencias Nucleares, Universidad Nacional Aut\'{o}noma de M\'{e}xico, 04510 Ciudad de M\'{e}xico, M\'{e}xico}

\author{A. Mart\'{i}n-Ruiz}
\email{alberto.martin@nucleares.unam.mx}
\affiliation{Instituto de Ciencias Nucleares, Universidad Nacional Aut\'{o}noma de M\'{e}xico, 04510 Ciudad de M\'{e}xico, M\'{e}xico}

\begin{abstract}
We theoretically investigate the Aharonov-Bohm effect in a thick-walled Weyl semimetal (WSM) cylinder subject to an external axial magnetic field. By employing a low-energy effective Hamiltonian, we analytically solve the eigenvalue problem for both infinite and finite-length cylindrical geometries. We apply infinite-mass boundary conditions at the radial walls and MIT bag boundary conditions at the cylinder caps to properly account for intra-node confinement and inter-valley scattering, respectively. Our numerical results demonstrate that the spatial separation of the Weyl nodes acts as an internal chiral gauge field. This geometric field intrinsically breaks time-reversal (TR) symmetry, lifting the chiral degeneracy even at zero external flux. This symmetry breaking manifests as spontaneous persistent currents and the unfolding of conductance channels. Furthermore, longitudinal confinement induces propagation-direction-dependent energy splitting, altering the partial density of states and causing spatio-chiral current imbalances.
\end{abstract}


\maketitle

\section{Introduction}

The discovery of topological phases of matter has profoundly reshaped the landscape of modern condensed matter physics, providing a fertile ground for realizing relativistic-like fermions in solid-state systems \cite{hasan2010colloquium,qi2011topological,bansil2016colloquium,chiu2016classification}. Following the successful theoretical prediction and experimental verification of two-dimensional graphene and three-dimensional topological insulators (TIs) \cite{castro2009electronic,moore2010birth,ando2013topological}, topological semimetals have emerged as a frontier of intensive research \cite{burkov2011topological,armitage2018weyl}. Among these, Weyl semimetals (WSMs) represent a significant conceptual leap, as they host three-dimensional massless excitations \cite{wan2011topological,weng2015weyl}. In a WSM, the bulk conduction and valence bands intersect at discrete, isolated points in the Brillouin zone, known as Weyl nodes \cite{xu2015discovery,lv2015experimental}. These nodes are protected by topology and necessarily come in pairs of opposite chirality, $\chi = \pm 1$, acting as monopoles and antimonopoles of Berry curvature \cite{nielsen1981absence,volovik2003universe,haldane2004berry}. This unique topological structure leads to a plethora of exotic transport phenomena, most notably the chiral magnetic effect, the chiral anomaly, and axionic electrodynamic responses \cite{fukushima2008chiral,son2012chiral,huang2015observation,zyuzin2012topological, perez6028257supersymmetric}. These topologically non-trivial effects manifest in macroscopic observables such as dramatic negative longitudinal magnetoresistance and the chiral Planar Hall effect \cite{perez2024dirac}. A hallmark of quantum coherence and macroscopic interference in mesoscopic systems is the Aharonov-Bohm (AB) effect \cite{aharonov1959significance}. 

Originally formulated for electrons in a vacuum, the AB effect demonstrates that an electromagnetic potential can measurably modify the quantum phase of a particle's wavefunction, even if the particle resides entirely in a region with zero electric and magnetic fields \cite{webb1985observation}. In the context of low-dimensional Dirac materials, the AB effect and related quantum interference phenomena has proven pivotal in unraveling the coherent charge and energy transport properties of carbon nanotubes, graphene rings, and disordered graphene landscapes \cite{bachtold1999aharonov,russo2008observation,3r17-kfy7,canas2026charge}. In these systems, threading a magnetic flux induces periodic quantum oscillations in the conductance and generates equilibrium, dissipationless persistent currents \cite{buttiker1983josephson,recher2007aharonov}. The extension of these principles to topological materials has unveiled even richer physics; for instance, in TIs, the AB effect serves as an unambiguous probe of surface states in nanowires, revealing clear distinctions between topologically non-trivial and trivial transport regimes due to the acquisition of a Berry phase \cite{peng2010aharonov,bardarson2010aharonov,zhang2010anomalous}. However, when considering WSMs, the interplay between external magnetic fluxes and the intrinsic topological properties of the material introduces a new paradigm of symmetry breaking. 

A key feature of a WSM is that the Weyl nodes are separated in momentum space by a vector $2\mathbf{b}$, or in energy by $2b_0$ \cite{halasz2012time,zyuzin2012weyl}. This separation inherently requires the breaking of either time-reversal (TR) symmetry, parity (P) symmetry, or both \cite{burkov2015chiral}. Remarkably, the spatial separation parameter $\mathbf{b}$ enters the low-energy effective Hamiltonian not merely as a shift in momentum, but effectively as an internal axial gauge vector potential \cite{grushin2012consequences,liu2013chiral,cortijo2015elastic}. Unlike an external electromagnetic vector potential $\mathbf{A}$, which couples symmetrically to the electric charge regardless of the valley (manifesting as the standard AB flux $\Phi$), the axial gauge field couples with opposite signs to electrons of opposite chirality \cite{pikulin2016chiral}. Consequently, $\nabla\times \mathbf{b}$ acts as an intrinsic chiral magnetic field that strongly competes or cooperates with the external AB flux \cite{ilan2020pseudo,baireuther2016scattering}. This geometric interplay provides a topological mechanism for the lifting of chiral degeneracy and intrinsically breaks TR symmetry, even in the absence of an external magnetic field \cite{berry1987neutrino}. 

In this work, we theoretically investigate the AB effect in a thick-walled WSM cylinder ($r_1 < r < r_2$). This specific geometry introduces a finite thickness $W = r_2 - r_1$, which allows for a more realistic description of experimental setups and reveals how radial confinement intricately affects the energy spectrum and persistent currents. By employing a full $4 \times 4$ low-energy effective Hamiltonian, we solve the eigenvalue problem for both infinite and finite-length cylindrical geometries pierced by an axial magnetic flux. To accurately capture the physics at the boundaries, we apply infinite-mass boundary conditions (IMBCs) at the radial walls, and MIT bag boundary conditions (BCs) at the longitudinal cylinder ends to properly account for inter-valley scattering. Our results demonstrate that the internal axial gauge field, driven by the Weyl node separation, breaks TR symmetry intrinsically, lifting the chiral degeneracy and inducing spontaneous persistent currents at zero flux. Furthermore, we show that longitudinal confinement in the finite cylinder introduces a propagation-direction-dependent energy splitting, which dramatically alters the partial density of states (PDOS) and induces complex spatial imbalances in the local chiral probability densities and currents. These findings provide clear observable signatures for flux-driven topological transport in mesoscopic WSM structures, offering a pathway to distinguish axial gauge fields from standard magnetic fields in confined geometries.

The remainder of this paper is organized as follows. In Sec. \ref{sect:model}, we present the low-energy effective Hamiltonian for a two-node WSM and formulate the eigenvalue problem in cylindrical coordinates, incorporating the AB flux via minimal coupling. Section \ref{sec:infinite} is devoted to the infinite-cylinder case, where we derive the energy spectrum under IMBCs at the radial walls and analyze the resulting persistent currents, conductance channels, and partial density of states. In Sec. \ref{sec:finite}, we extend the analysis to a finite-length cylinder, introducing MIT bag boundary conditions at the longitudinal caps to account for inter-valley scattering, and we examine how longitudinal confinement modifies the energy spectrum and the spatio-chiral structure of probability densities and currents. Finally, Sec. \ref{sec:conclusions} summarizes our main findings and discusses their experimental implications for mesoscopic WSM structures.

\section{Model}
\label{sect:model}

We start from the full $4\times 4$ effective low-energy Hamiltonian describing a WSM with two-nodes of opposite chirality,
\begin{equation}\label{eq:Hamiltonian44}
    H(\mathbf{k})= b_0 \tau_z \otimes \sigma_0 + \hbar v_F \left[\tau_z \otimes (\boldsymbol{\sigma}\cdot \mathbf{k}) -\tau_0 \otimes (\boldsymbol{\sigma}\cdot \mathbf{b})\right],
\end{equation}
where $b_0$, $\mathbf{b}$ represent the temporal (energy) and spatial (momentum) separation of the two chiral cones in the WSM, $\hbar$ is the reduced Planck's constant, $v_F$ denotes the Fermi velocity in the material, and $\mathbf{k}= \mathbf{p}/\hbar$ is the wave vector. In Eq. (\ref{eq:Hamiltonian44}), $\boldsymbol{\tau}=(\tau_x,\tau_y,\tau_z)$ are Pauli matrices acting on the chirality (node) subspace (+,-), whereas $\boldsymbol{\sigma}=(\sigma_x,\sigma_y,\sigma_z)$ are Pauli matrices acting on the spin degree of freedom within each Weyl cone. The contribution $b_0\tau_z$ reproduces the chirality-dependent energy shift of the two Weyl nodes. The kinetic term $\tau_z\otimes (\boldsymbol{\sigma}\cdot \mathbf{k})$ correctly implements the handedness of both cones. Finally, the term $-\tau_0\otimes (\boldsymbol{\sigma}\cdot \mathbf{b})$ shows that the vector $\mathbf{b}$ enters with the same sign at both chiral sectors. Hamiltonian in (\ref{eq:Hamiltonian44}) is block-diagonal in the chirality basis, i.e.,
\begin{equation}\label{eq:block}
    H(\mathbf{k}) = \begin{pmatrix}
        H_+(\mathbf{k}) & 0\\
        0 & H_-(\mathbf{k})
    \end{pmatrix},
\end{equation}
 and acts over the four-component chiral spinor
 \begin{equation}
     \Psi(\mathbf{k}) = \begin{pmatrix}
         \Psi_+(\mathbf{k})\\\Psi_-(\mathbf{k})
     \end{pmatrix},
 \end{equation}
where the subscripts $\pm$ refer to right/left-handed chirality, respectively. Consequently, the eigenvalue equation $H(\mathbf{k})\Psi(\mathbf{k})=E\Psi(\mathbf{k})$ describes two decoupled Weyl fermions at low-energy theory.
 
Hamiltonian in (\ref{eq:Hamiltonian44}) has the possibility of having a system with broken TR and P symmetries at the same time, however, for simplicity, we are interested in the case in which the Weyl nodes are spatially-separated in $\hat{z}$-direction, i.e. $\bm{b} = b \hat{z}$ \footnote{Other directions can be treated analogously, but break cylindrical symmetry.}, and also $b_{0\chi}=0$, so that preserving P symmetry while breaking PT symmetry. In this case, the chiral-Hamiltonians in (\ref{eq:block}) can be written as
\begin{align}\label{eq:Hamiltonian_independent}
H_{\chi}(\mathbf{k}) &= \chi \hbar v_F \boldsymbol{\sigma} \cdot (\mathbf{k} - \chi b\hat{z}) \nonumber \\
&= \chi \hbar v_F 
    \begin{pmatrix} 
        -i\partial_z -\chi b & -i\partial_x - \partial_y \\ \\
        -i\partial_x + \partial_y & i\partial_z +\chi b 
    \end{pmatrix} ,
\end{align}
where $\chi=\pm 1$ represents the chirality, and where we have made the substitution $k_j = -i\partial_j \equiv -i \frac{\partial}{\partial j}$, $j=x,y,z$.

\begin{figure}[htbp]
  \centering
    \includegraphics[width=0.8\columnwidth]{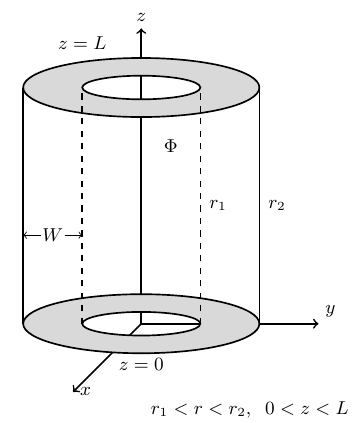}

  \caption{Thick-walled WSM cylinder with internal radius $r_1$ and external radius $r_2$, with longitudinal axis in $z$-direction. In this figure the cylinder is also finite in the longitudinal direction, and a magnetic flux $\Phi$ is considered in the inner cavity of the cylinder, but a null magnetic field inside the WSM is considered (AB setup).}
  \label{fig:cilindroWSM}
\end{figure}

We are interested in studying the AB effect \cite{aharonov1959significance} inside the WSM thick-walled cylinder of inner radius $r_1$ and outer radius $r_2$, for the case of infinite length ($-\infty < z <\infty$), and when we confine the cylinder in the longitudinal direction as well ($0 < z <L$), see Fig. \ref{fig:cilindroWSM} for later case. In this sense, we consider a vector potential coupled to the charge carriers in the material through minimal coupling $\mathbf{k}\rightarrow\mathbf{k}+\frac{e\mathbf{A}}{\hbar}$, depending in $r$ of the form 
\begin{equation}
    \mathbf{A}(r)= \frac{\Phi}{2\pi r} \mathbf{e}_{\theta},
\end{equation}
which produces a total magnetic flux $\Phi$ through the $z$-axis, while keeping $\mathbf{B}=0$ inside the WSM. 

Because of the symmetry considered, it is convenient to use cylindrical coordinates ($r,\theta,z$). The Cartesian derivatives become
\begin{align*}
\partial_x = \cos\theta \partial_r - \frac{\sin\theta}{r} \partial_\theta, \qquad
\partial_y = \sin\theta \partial_r + \frac{\cos\theta}{r} \partial_\theta, 
\end{align*}
where $\theta=\text{arctan}(y/x)$. Under these assumptions, the Hamiltonian in (\ref{eq:Hamiltonian_independent}) takes the form 
\begin{align}\label{eq:Hamiltonian_Cylindric}
H_{\chi}(\mathbf{k})
&= \chi \hbar v_F \begin{pmatrix}
-i\partial_z - \chi b & e^{-i\theta}\left[-i\partial_r - \frac{1}{r}\partial'_\theta\right] \\ \\
e^{i\theta}\left[-i\partial_r + \frac{1}{r}\partial'_\theta\right] & i\partial_z +\chi b
\end{pmatrix},
\end{align}
where $\partial'_\theta = \partial_\theta + i\frac{\Phi}{\Phi_0}$ includes the magnetic flux, and where we have defined the magnetic flux quantum as $\Phi_0=2\pi\hbar/e$. 

Let us now exploit the fact that the Hamiltonian (\ref{eq:Hamiltonian_Cylindric}) commutes with the total angular momentum operator, $J_z= L_z + \hbar \sigma_z/2$, where $L_z=-i\partial_\theta$, so that both operators share the same basis. Explicitly, 
\begin{equation}
L_z \Psi_{\chi} = -i\partial_\theta \Psi_{\chi} = \begin{pmatrix} j-\frac{1}{2}& 0 \\ \\ 0 & j+\frac{1}{2} \end{pmatrix} \Psi_{\chi}.
\end{equation}
We propose the ansatz
\begin{equation}
\Psi_{\chi}(r,\theta,z) = e^{i(j-1/2)\theta} \begin{pmatrix} \varphi_1^{\chi}(r,z) \\ \varphi_2^{\chi}(r,z)e^{i\theta} \end{pmatrix},
\end{equation}
which results in the eigenvalue equation 
\begin{equation}\label{eq:eigenequation}
H_{\chi}\varphi^{\chi}(r,z) = E \varphi^{\chi}(r,z),  
\end{equation}
with the new spinor 
\begin{equation}
    \varphi^{\chi}(r,z) = (\varphi_1^{\chi}(r,z),\ \varphi_2^{\chi}(r,z))^T,
\end{equation} 
where we define the spinor associated to the Weyl cones with right and left chirality, respectively. The Hamiltonian takes the form
\begin{equation}\label{eq:Dirac}
H_{\chi} = \chi \hbar v_F\begin{pmatrix} -i\partial_z-\chi b & -i[\partial_r + \frac{1}{r}(j'+\frac{1}{2})] \\ \\ -i[\partial_r - \frac{1}{r}(j'-\frac{1}{2})] & i\partial_z+\chi b \end{pmatrix},
\end{equation}
with $j'=j+\Phi/\Phi_0$. We see that the magnetic flux shifts the angular momentum quantum number by $\Phi/\Phi_0$, which is the hallmark of the AB effect.

In the next section, we solve the eigenvalue equation for the infinite cylinder case, applying IMBCs \cite{berry1987neutrino, stockmeyer2019} in $r$-borders. Subsequently, in Sec. \ref{sec:finite}, we study the case of a finite cylinder (Fig. \ref{fig:cilindroWSM}) applying additional MIT bag BCs over the $z$ edges. In both cases we analyze the resulting spectrum and its dependence over the dimensionless magnetic flux $\Phi/\Phi_0$.

\section{Infinite cylinder}
\label{sec:infinite}

When the cylinder is infinite in $z$-direction, and as $\partial_z$ commutes with $H_{\chi}(\mathbf{k})$, then one can propose the ansatz
\begin{equation}
    \varphi_{1,2}^\chi (r,z) = e^{ik_z z} R^{\chi}_{1,2}(r). 
\end{equation}
By defining $k_z^\chi = k_z-\chi b$, the inverse-length spectral parameter $\lambda = E/\hbar v_F$ \footnote{$\lambda$ possesses units of inverse length, consistent with the derivatives derivatives and $b$ appearing in the Hamiltonian $H_\chi(\mathbf{k})$.}, and the radial operators
\begin{equation}
    \hat{D}_r^{\pm} = \left[\partial_r \pm \frac{1}{r}(j'\pm \frac{1}{2})\right],
\end{equation}
the eigenvalue equation (\ref{eq:eigenequation}) turns into the pair of coupled linear equations (for each chirality) presented next
\begin{equation}
\left\lbrace
\begin{aligned}\label{eq:coupled}
\hat{D}_r^+ \varphi^{\chi}_2 &= i \left[\chi \lambda -k_z^\chi\right] \varphi^{\chi}_1 \\
\hat{D}_r^- \varphi^{\chi}_1 &= i \left[\chi \lambda +k_z^\chi\right] \varphi^{\chi}_2
\end{aligned}    
\right. .
\end{equation}

The operators $\hat{D}_r^\pm$ posses the following properties:
\begin{itemize}
    \item $\hat{D}_r^+ \hat{D}_r^-= \partial_r^2+ \frac{1}{r}\partial_r-\frac{1}{r^2}(j'-\frac{1}{2})^2$.
    \item $\hat{D}_r^- \hat{D}_r^+= \partial_r^2+ \frac{1}{r}\partial_r-\frac{1}{r^2}(j'+\frac{1}{2})^2$.
    \item $[\hat{D}_r^+,\hat{D}_r^-]= 1/r^2$.
\end{itemize}
With these properties it is easy to decouple Eqs. (\ref{eq:coupled}), resulting in two Bessel differential equations for the radial functions of the form
\begin{align}\label{eq:Bessel1}
\left[\partial_r^2 + \frac{1}{r}\partial_r - \frac{1}{r^2}\left(j'\mp\frac{1}{2}\right)^2 + \alpha_{\chi}^2 \right]R^{\chi}_{1,2} =0,
\end{align}
where $\alpha_{\chi} = \sqrt{\lambda^2 -(k_z - \chi b)^2}$. Equation (\ref{eq:Bessel1}) has as general solution a linear combination of $J_\nu(r)$ and $Y_\nu(r)$, Bessel functions of the first and second type, respectively, as follows
\begin{equation}
R_{1,2}^{\chi}(r) = A^\chi_{1,2} J_{\nu_{\mp}}(\alpha_{\chi} r) + B^\chi_{1,2} Y_{\nu_{\mp}}(\alpha_{\chi} r),
\end{equation}
where $\nu_{\mp}=j' \mp\frac{1}{2}$, $\nu_+=\nu_-+1$. Through the first-order equations in (\ref{eq:coupled}), and using the standard Bessel identities
\begin{equation}\label{eq:Bessel_Id1}
\left(\partial_r + \frac{j+1/2}{r}\right) J_{j+1/2}(\alpha r) = \alpha J_{j-1/2}(\alpha r),
\end{equation}
\begin{equation}\label{eq:Bessel_Id2}
\left(\partial_r - \frac{j-1/2}{r}\right) J_{j-1/2}(\alpha r) = -\alpha J_{j+1/2}(\alpha r),
\end{equation}
and analogously for the $Y_{j\pm 1/2}(\alpha r)$ functions, it is possible to relate the constants \begin{equation}
    A^\chi_2= i\beta_\chi A^\chi_1, \qquad B^\chi_2= i\beta_\chi B^\chi_1,
\end{equation} 
where we have defined 
\begin{equation}
    \beta_\chi = \frac{\chi\lambda-(k_z-\chi b)}{\alpha_\chi}.
\end{equation}
It is worth noting that for $k_z=0$, $b=0$, that is confining the movement to the ($r, \theta$) plane with no node separation, we recover the factor for graphene $\beta_\chi=\text{sgn}(E)$. With these relations, the radial functions are written as
\begin{equation}
R_{1}^{\chi}(r) = A^\chi_1 J_{\nu_-}(\alpha_{\chi} r) + B^\chi_1 Y_{\nu_-}(\alpha_{\chi} r),
\end{equation}
\begin{equation}
R_2^{\chi}(r) = i \beta_\chi [A^\chi_1 J_{\nu_+}(\alpha_{\chi} r) + B^\chi_1 Y_{\nu_+}(\alpha_{\chi} r)].
\end{equation}

\subsection{Energy spectrum}\label{subsec:energy_inf}

The energy spectrum of the system, as well as a relation between $A^\chi_1$ and $B^\chi_1$ comes from the IMBCs in the radial inner $r=r_1$ and outer $r=r_2$ borders. The IMBCs we consider here are applied independently over each Weyl node, that is, we are considering a smooth BC in which $r_1,r_2\gg a$, with $a$ the lattice constant. For this approximation the corresponding intra-node BCs are given by 
\begin{equation}\label{eq:BCs}
    \Psi_\chi(r,\theta,z)= i\chi(\boldsymbol{\sigma}\cdot \mathbf{n})\Psi_\chi(r,\theta,z),
\end{equation}
where $\mathbf{n}$ is the vector normal to the surface of the cylinder, defined by $\mathbf{n}=\hat{r}=\pm(\cos\theta, \sin \theta)$, where the plus/minus sign refers to the outer/inner boundary, respectively. So that the BC in (\ref{eq:BCs}) turns into
\begin{equation}\label{eq:r_conditions}
    \Psi_\chi(r,\theta,z)= \pm i\chi\sigma_r\Psi_\chi(r,\theta,z),
\end{equation}
where $\sigma_r=\begin{pmatrix}
    0 & e^{-i\theta}\\ 
    e^{i\theta} & 0
\end{pmatrix}$.

In the inner radius $r=r_1$:
\begin{equation}
R_1^\chi = -i \chi R_2^\chi \bigg|_{r=r_1},
\end{equation}
which yields
\begin{align}
& A_1^\chi J_{\nu_-}(\alpha_\chi r_1) + B_1^\chi Y_{\nu_-}(\alpha_\chi r_1) \nonumber \\ &=  (-i\chi) i\beta_{\chi} \left[ A_1^\chi J_{\nu_+}(\alpha_\chi r_1) + B_1^\chi Y_{\nu_+}(\alpha_\chi r_1) \right].
\end{align}
Rearranging terms we have
\begin{align}
	& A_1^\chi [J_{\nu_-}(\alpha_\chi r_1) - \chi\beta_{\chi} J_{\nu_+}(\alpha_\chi r_1)] \nonumber \\ &+ B_1^\chi [Y_{\nu_-}(\alpha_\chi r_1) - \chi\beta_{\chi} Y_{\nu_+}(\alpha_\chi r_1)] = 0 .
\end{align}

Proceeding in a similar way for the outer radius $r=r_2$: 
\begin{equation}
    R_1^\chi = i \chi R_2^\chi \big|_{r=r_2}
\end{equation}
reads
\begin{align}
	& A_1^\chi [J_{\nu_-}(\alpha_\chi r_2) +\chi\beta_{\chi} J_{\nu_+}(\alpha_\chi r_2)] \nonumber \\ &+ B_1^\chi [Y_{\nu_-}(\alpha_\chi r_2) +\chi \beta_{\chi} Y_{\nu_+}(\alpha_\chi r_2)] = 0 .
\end{align}

\begin{figure}
    \centering
    \includegraphics[width=0.9\columnwidth]{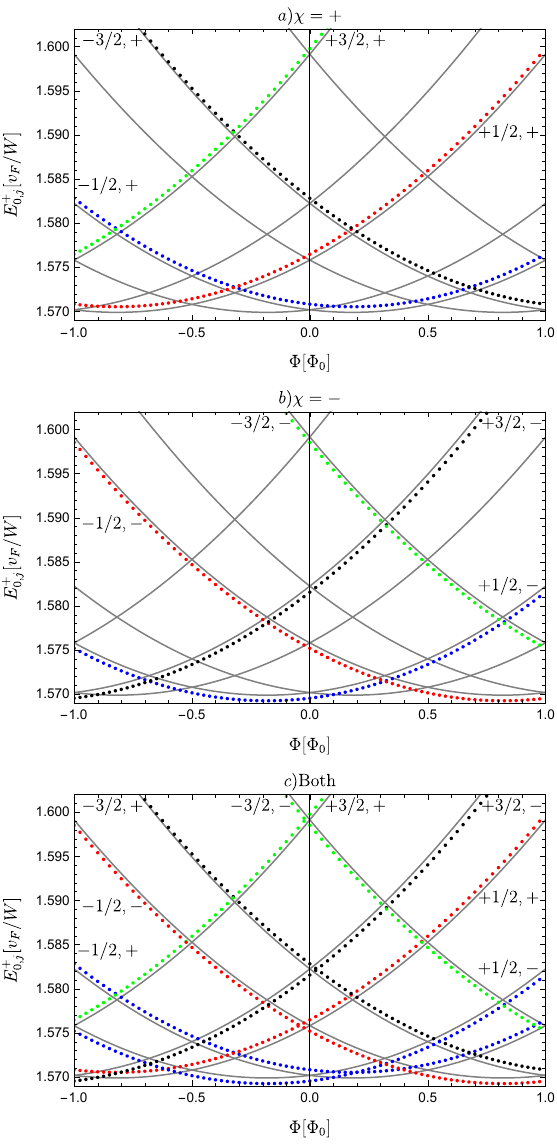}
    \caption{Energy eigenvalues as a function of the magnetic flux, $\Phi$, for different values of angular momentum, $j$, for the case $b=0.001$, $k_z=0$. Energy eigenvalues for states with positive and negative chirality are shown in $a$) and $b$), respectively. In c) energies for states with both chiralities are presented. In all the figures in the paper we have used the values $r_1=5.5$, $r_2=6.5$.}
    \label{fig:infinite_energy_k0_bn0}
\end{figure}

Both BCs can be represented as a matrix equation 
\begin{equation}\label{eq:matrix_eq}
    \begin{pmatrix}
        M_{11} & M_{12}\\ M_{21} & M_{22}
    \end{pmatrix}\begin{pmatrix}
        A^\chi_1 \\ B^\chi_1
    \end{pmatrix} = \begin{pmatrix}
        0 \\ 0
    \end{pmatrix},
\end{equation}
where
\begin{align*}
    M_{11}&= J_{\nu_-}(\alpha_\chi r_1) - \chi\beta_\chi J_{\nu_+}(\alpha_\chi r_1),\nonumber \\
    M_{12}&= Y_{\nu_-}(\alpha_\chi r_1) - \chi\beta_\chi Y_{\nu_+}(\alpha_\chi r_1),\nonumber \\
    M_{21}&= J_{\nu_-}(\alpha_\chi r_2) + \chi\beta_\chi J_{\nu_+}(\alpha_\chi r_2),\nonumber \\
    M_{22}&= Y_{\nu_-}(\alpha_\chi r_2) + \chi\beta_\chi Y_{\nu_+}(\alpha_\chi r_2).
\end{align*}
We define the matrix $\mathbb{M}$ as
\begin{equation}\label{eq:matrix}
    \mathbb{M} = \begin{pmatrix}
        M_{11} & M_{12}\\
        M_{21} & M_{22}
    \end{pmatrix}.
\end{equation}
In order to find a nontrivial solution of Eq. (\ref{eq:matrix_eq}), the condition $\det (\mathbb{M})=0$ must be satisfied. This relation gives a transcendental equation for the energy eigenvalues which depends on $\chi$.

\begin{figure}
    \centering
    \includegraphics[width=0.9\columnwidth]{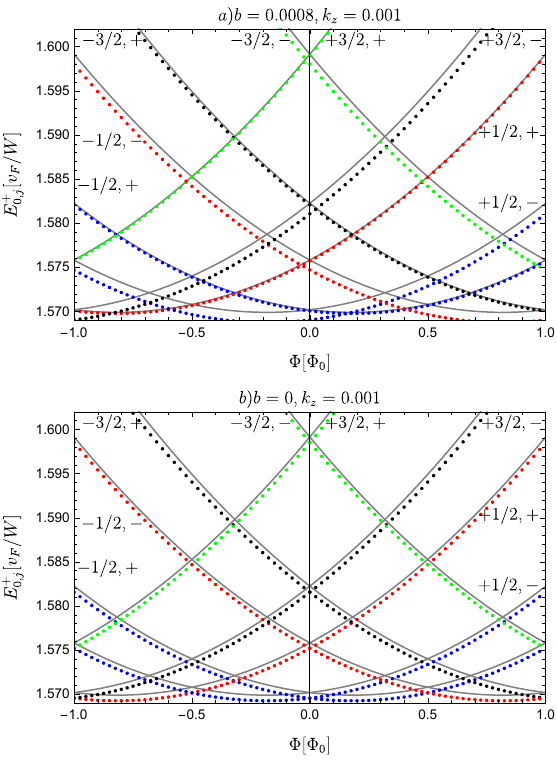}
    \caption{Energy eigenvalues as a function of the magnetic flux, $\Phi$, for different values of angular momentum, $j$. $a$) for the case $b=0.0008$, $k_z=0.001$. $b$) for the case $b=0$, $k_z=0.001$. In both cases, energies for states with both chiralities are presented.}
    \label{fig:infinite_energy_kn0_bn0}
\end{figure}
In our numerical calculations, we set $\hbar=e=1$ and use a characteristic length scale $l_0\equiv 15\,\mathrm{nm}$ to define dimensionless variables. In this convention, $r_1=5.5$, $r_2=6.5$, and $L=100$ denote lengths measured in units of $l_0$, i.e., the corresponding physical lengths are $r_1l_0$, $r_2l_0$, and $Ll_0$. The parameters $b$ and $k_z$ are likewise measured in units of $l_0^{-1}$. This maps our theoretical geometry to a physical topological cylinder with an outer radius of $r_2 \sim 100$ nm and a length of $L \sim 1.5\ \mu\text{m}$. At these realistic dimensions, an external magnetic field of roughly $\sim 0.2$ T is sufficient to thread a fundamental flux quantum $\Phi_0$ through the inner cavity. 

Figure \ref{fig:infinite_energy_k0_bn0} depicts some of the lowest energy eigenvalues for $n=0$, and $j=\pm 1/2, \pm 3/2$, as a function of the magnetic flux $\Phi$, where the values $b=0.001$ and $k_z=0$ where taken. Dots in this figure represent the corresponding energy values computed directly from the condition $\det (\mathbb{M})=0$, while the thin gray lines represent the energy spectrum for the graphene case ($k_z=0$, $b=0$) taken here as a reference. Figure \ref{fig:infinite_energy_k0_bn0}$a$) depicts the energy levels for positive ($\chi=+$) chirality, showing a positive shift in energies compared to the graphene case (gray lines). On the other hand, from Fig. \ref{fig:infinite_energy_k0_bn0}$b$) we can note a negative energy shift of the energy levels for negative ($\chi=-$). Figure \ref{fig:infinite_energy_k0_bn0}$c$) shows the energy levels for both chiralities.  From this figure we can observe that the separation of the chiral nodes in WSMs causes an asymmetry in the total energy spectrum (considering the two chiralities), i.e. $E_{0,j}^+\neq E_{0,-j}^{-}$, resulting from the positive/negative shifting of energies with positive/negative chirality. An important consequence of this behavior is that TR symmetry is broken even when no net magnetic flux is present, differently from the graphene case which conserves TR symmetry. As a consequence, chiral degeneracy for $\Phi=0$ is lifted in WSMs. 

When we provide the particles a momentum perpendicular to the ($r,\theta$) plane, i.e. $k_z\neq 0$, we observe a negative shifting of energies with the same magnitude for the two chiralities, as shown in Fig. \ref{fig:infinite_energy_kn0_bn0}. Therefore, when a chiral node separation exists ($b\neq 0$) the energy asymmetry, and thus the TR symmetry breaking remains for particles moving on the whole infinite cylinder. On the other hand, when no node separation is considered ($b=0$), TR symmetry is recovered (chiral degeneracy). This behavior is shown in Figs. \ref{fig:infinite_energy_kn0_bn0}$a$) and \ref{fig:infinite_energy_kn0_bn0}$b$), respectively, where the values $b=0.0008, k_z=0.001$ and $b=0, k_z=0.001$ where taken in each case.

To gain a deeper physical insight into the symmetry breaking mechanism observed in the spectrum, it is instructive to recognize that the chiral node separation parameter, $b$, enters the effective Hamiltonian effectively as an internal axial gauge field. Unlike the external vector potential $A_\theta$, which couples symmetrically to the electric charge regardless of the valley (manifesting as the macroscopic AB flux $\Phi$), the axial gauge field couples with opposite signs to states of opposite chirality, $\chi$. Consequently, $b$ and the external AB flux strongly compete or cooperate depending on the specific valley index. This interplay elegantly explains why the chiral degeneracy is intrinsically lifted. More importantly, it provides a topological origin for the non-vanishing spontaneous persistent currents at zero external flux ($\Phi=0$). The system intrinsically harbors a circulating chiral current driven exclusively by this internal gauge field, which macroscopically manifests as a valley polarization.

\begin{figure*}[ht]
    \centering
    \includegraphics[width=0.99\textwidth]{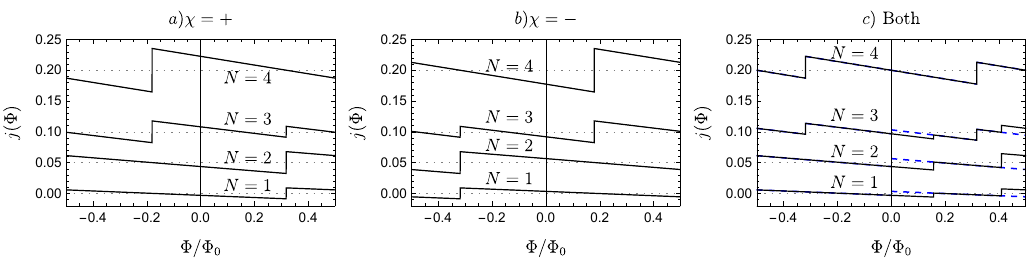}
    \caption{Persistent current for the case $b=0.0008$, $k_z=0.001$ [Fig. \ref{fig:infinite_energy_kn0_bn0}$a$)]. $a$) Shows the persistent current for positive chirality, $b$) for negative chirality, whereas in $c$) the sum of currents for both chiralities is shown.}
    \label{fig:infinite_persistent}
\end{figure*}

In the next subsections we study the consequences of the TR symmetry breaking caused by the node separation in WSMs in some observables such as the persistent current, the conductance channels and the charge probability densities and currents in the system. We also study the PDOS considering magnetic flux in the system.

\subsection{Persistent current inside the cylinder}
In mesoscopic physics, a persistent current is an equilibrium, perpetual, and dissipationless electrical current induced by a magnetic flux, even in the absence of an applied voltage. It is a direct macroscopic manifestation of the AB effect and of quantum phase coherence in condensed matter systems. By applying a fractional magnetic flux, electrons moving clockwise and counterclockwise no longer have the same energy (TR symmetry is broken). Mathematically, the persistent current in the infinite thick-walled cylinder we are studying is given by
\begin{equation}
    j(\Phi) = - \sum_{\chi=\pm} \sum_{n,m} \frac{\partial E_{nm}^\chi}{\partial \Phi},
\end{equation}
where the second sum runs over all occupied states. 

In Fig. \ref{fig:infinite_persistent} we show the persistent current as a function of the magnetic flux for different number of electrons in the thick-walled cylinder, $N=1,2,3,4$ for states with $E>0$ for the system under study for the values $b=0.0008$ and $k_z=0.001$, corresponding to the energy spectrum given in Fig. \ref{fig:infinite_energy_kn0_bn0}$a$)  (black solid lines). For comparison purposes, have included the persistent current for the graphene case, $b=0$, $k_z=0$ (dashed blue lines). Figures \ref{fig:infinite_persistent}$a)$, $b$) show the persistent current for particles with positive chirality $\chi=+$, $\chi=-$, respectively. We can observe no changes in the persistent current compared with the graphene system, presenting a finite persistent current (valley polarization) at $\phi=0$ as shown in Ref. \cite{recher2007aharonov}. This is because, as seen in Sec. \ref{subsec:energy_inf}, the spectrum for each chirality just shifts upward or downward for nonzero values of $b$ and $k_z$. Figure \ref{fig:infinite_persistent}$a)$ shows the persistent current considering both chiralities. In this case, as the spectrum for positive chirality shifts upward while for negative chirality shifts downward, the energy bands for different chiralities stop crossing at $\Phi=0$, deviating the crossing points to positive and negative values of $\Phi$, as seen in Fig. \ref{fig:infinite_energy_kn0_bn0}$a$). This deviation of the crossings generates new substructures (kinks) in the persistent current in comparison to the graphene case, and generates a finite persistent current at $\Phi=0$, that is, the system presents a finite persistent current even without magnetic flux, exclusively induced by the separation of the chiral nodes.

\begin{figure}
    \centering
    \includegraphics[width=0.9\columnwidth]{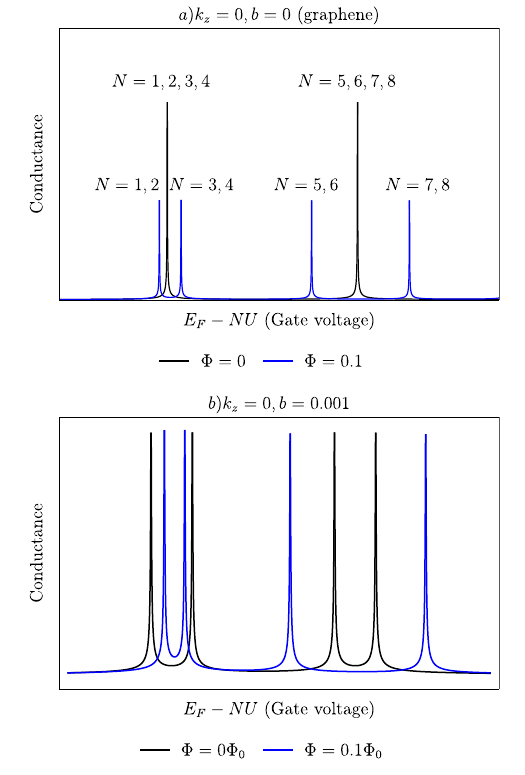}
    \caption{Conductance resonances (conductance channels) through the infinite thick-walled cylinder coupled to leads as a function of the Fermi energy (gate voltage), for $\Phi=0$ (black lines) and $\Phi=0.1$ (blue lines). $a$) Shows the graphene case, $b=0$, $k_z=0$. $b$) $b=0.001$, $k_z=0$.}
    \label{fig:conductances}
\end{figure}

\begin{figure}
    \centering
    \includegraphics[width=0.95\columnwidth]{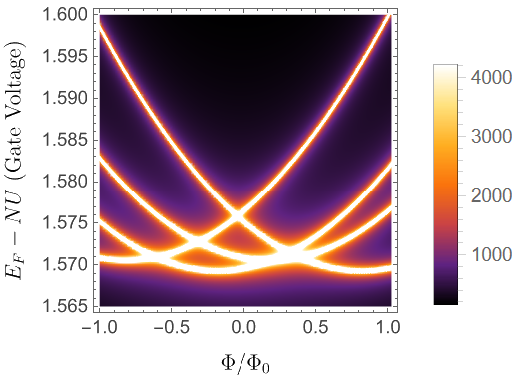}
    \caption{Density plot of the conductance resonances (conductance channels) through the thick-walled cylinder coupled to leads as a function of the Fermi energy (gate voltage), for $b=0.001$, $k_z=0$.}
    \label{fig:infinite_cond_density}
\end{figure}

\subsection{Conductance channels}

Considering a constant interaction model \cite{sohn2013mesoscopic} with charging energy U \footnote{Following standard electrostatic capacitance models for 3D topological cylinders, the charging energy for a cylinder of these scaled dimensions over a typical SiO$_2$ dielectric gate is estimated to be $U \approx 2-4$ meV. This scale is highly consistent with recent experimental measurements of Coulomb blockade and quantum dot formation in topological Dirac semimetal nanowires of comparable dimensions \cite{jung2018quantum}.}, in Fig. \ref{fig:conductances} we computed the conductance resonances (conductance channels) through the thick-walled cylinder coupled to leads as a function of the Fermi energy (gate voltage). Figure \ref{fig:conductances}$a$) represents the graphene case ($b=0$, $k_z=0$) studied in \cite{recher2007aharonov}, where at $\Phi=0$ the conductance presents a quadruple degeneration coming from spin and valley degeneracies (black lines); this fourfold degeneracy is broken into two twofold degenerated conductance channels (blue lines) as valley degeneracy is broken. Nevertheless, when a separation of chiral nodes is present, the system naturally breaks chiral degeneracy at $\Phi=0$, and consequently fourfold degenerated conductances are decoupled into two twofold degenerated conductance channels, as shown in Fig. \ref{fig:conductances}$b$). A density plot of the conductance resonances for different values of the magnetic flux and gate voltage is presented in Fig. \ref{fig:infinite_cond_density}, perfectly recovering the energy spectrum of Fig. \ref{fig:infinite_energy_k0_bn0}$c$).

\begin{figure*}[ht]
    \centering
\includegraphics[width=0.97\textwidth]{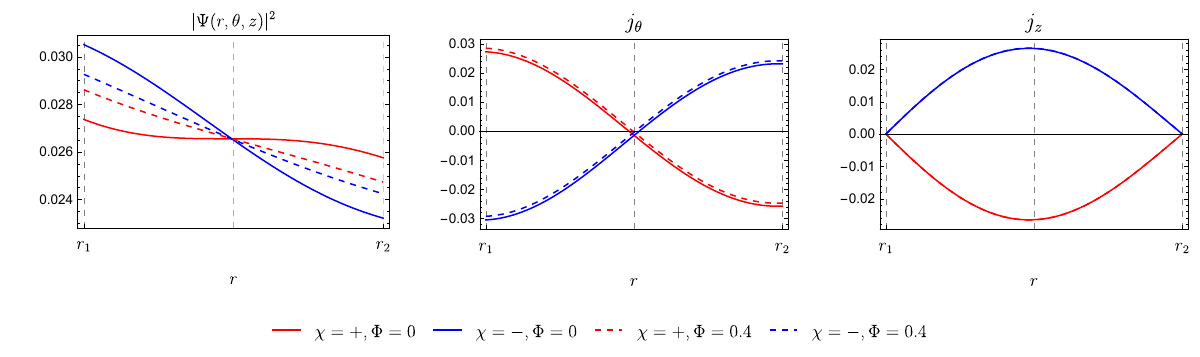}
    \caption{Probability density and currents for particles with defined chirality $\chi=+1$ (red lines) and $\chi=-1$ (blue lines) for the infinite cylinder problem. Solid lines refer to the case with no magnetic flux ($\Phi=0$) in the cylinder, while dashed lines correspond to finite magnetic flux ($\Phi=0.4$).}  
    \label{fig:infinite_currents}
\end{figure*}

\subsection{Density probabilities and currents}

In order to gain insight of the distribution and motion of the particles in the system, we calculate the corresponding probability density and probability currents for each chirality. They are defined, in the chiral representation, as
\begin{equation}
    \rho_\chi = \Psi^{\dagger}_\chi(r,\theta,z)\Psi_\chi(r,\theta,z),
\end{equation}
\begin{equation}
    j^i_\chi = \chi\Psi^{\dagger}_\chi(r,\theta,z)\sigma_i\Psi_\chi(r,\theta,z),\quad i=r,\theta,z,
\end{equation}
where $\sigma_i$ are the Pauli matrices in cylindrical coordinates. Figure \ref{fig:infinite_currents} shows the probability density and current densities for both chiralities (red lines for $\chi=+$ and blue lines for $\chi=-$, respectively) and for zero and finite magnetic flux (solid lines for $\Phi=0$ and dashed lines for $\Phi=0.4$, respectively). The density probability shows a concentration of chiral particles around the internal ($r_1$) radius, but with different magnitude for each chirality, while a chiral density balance is present near the center of the thick-walled cylinder. Also, at the presence of a magnetic flux, the imbalance between chiral densities is reduced, tending to present the same behavior for large values of $\Phi$. On the other hand, the angular current $j_{\theta}$ shows a counter-propagating behavior for particles with different chirality, which is a fingerprint of the AB effect, however, an imbalance in angular currents is present even at $\Phi=0$, as an effect of the node separation $b\neq 0$, and the effect of the magnetic flux is to move the currents upward/downward for positive/negative values of $\Phi$. Finally, the current in $z$ direction vanishes at the internal and external radii of the thick-walled cylinder, showing a parabolic growth while approaching to the center; we note no changes in $j_z$ when a magnetic flux is included. In this infinite case, the probability density and current densities remain constant for all $z \in (-\infty,\infty)$ and $\theta \in [0,2\pi]$ for $r$ fixed.

\subsection{Density of states}

To accurately evaluate the electronic density of states (DOS) in a confined WSM, particularly within a cylindrical geometry, a direct summation over an infinite set of discrete radial quantum numbers often introduces severe regularization challenges and arbitrary energy cutoffs. Instead of attempting to diagonalize the Hamiltonian in the full confined volume, a mathematically elegant and physically transparent approach is to employ the Krein-Friedel-Lloyd scattering formula \cite{lloyd1967wave,faulkner1980calculating}. Originally developed within the framework of multiple scattering theory, the Lloyd formula expresses the variation in the DOS, $\Delta \rho(E)$, as the energy derivative of the trace of the logarithm of the matrix $\mathbb{M}(E)$:
\begin{equation}
\Delta \rho(E) = \frac{1}{2\pi i} \frac{d}{dE} \text{Tr} \ln \mathbb{M}(E),
\end{equation}
where $\mathbb{M}(E)$ is the characteristic matrix from which the energy dispersion relation of the system is obtained, Eq. (\ref{eq:matrix}). In the context of our cylindrical WSM, this formalism allows us to treat the physical boundaries of the cylinder as scattering walls. The energy derivative of the trace-log term provides a clean, regularized spectral density, which is the essential ingredient for calculating the flux-induced quantum oscillations and the macroscopic persistent currents.

\begin{figure*}
    \centering    \includegraphics[width=1.\textwidth]{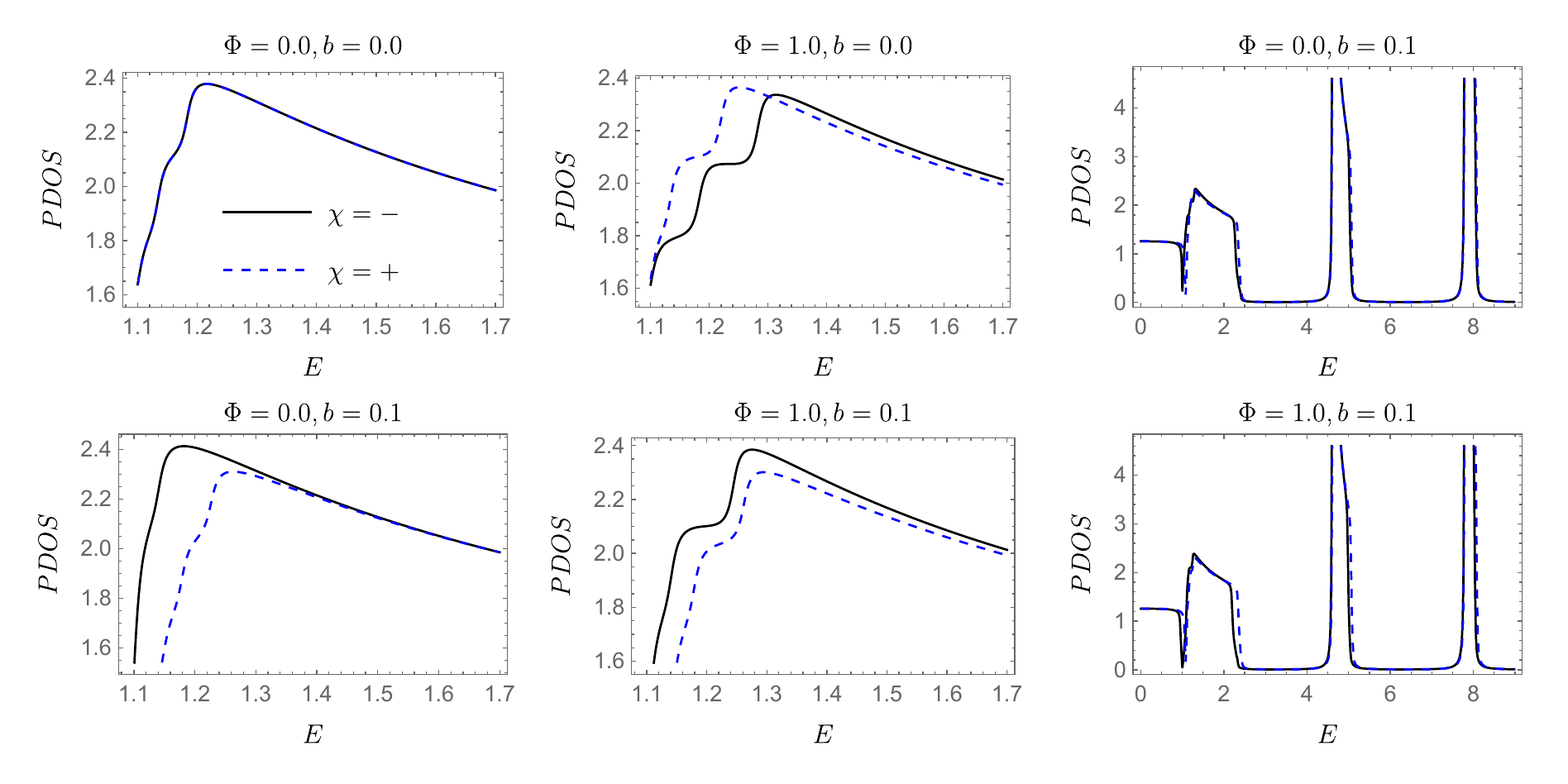}
    \caption{PDOS for the infinite cylinder for different values of the magnetic flux $\Phi$ and the separation of the chiral nodes $b$. Black solid lines represent the case of $\chi=-$, while dotted, blue lines represent the case of $\chi=+$.}
    \label{fig:infinite_DOS}
\end{figure*}

Figure \ref{fig:infinite_DOS} shows the calculated PDOS for each plus/minus chirality of the infinite cylinder (black and dashed blue lines, respectively). When we have the graphene case, $b=0$, $k_z=0$, we note no differences between the PDOS for each chirality, that is, they are degenerate (as previously discussed). We can also see that the effect of  a finite node separation, $b\neq 0$, is to shift energy states to lower values for $\chi=-$ while shifting to greater values for $\chi=+$, thus unfolding the PDOS for each chirality. On the other hand, positive values of the magnetic flux, $\Phi$, shift energy states to greater values for $\chi=-$ while shifting to lower values for $\chi=+$, and conversely for negative magnetic flux, also unfolding the chiral PDOS. In conclusion, the system presents two mechanisms to break TR symmetry, one by means of a node separation of the chiral nodes, and other by the effects of an external magnetic flux (AB effect).

\section{Finite case}
\label{sec:finite}
Let us now consider a thick-walled cylinder finite in $z$-direction ($0\leq z \leq L$). In this case the eigenvalue equation (\ref{eq:eigenequation}) leads to the next system of coupled linear differential equations
\begin{equation}
\left\lbrace
\begin{aligned}\label{eq:decoupled_finite}
\hat{D}_r^- \hat{D}_r^+ \varphi_{1}^\chi(r,z) = \hat{D}_z^- \hat{D}_z^+ \varphi_{1}^\chi(r,z) \\
\hat{D}_r^+ \hat{D}_r^- \varphi_{2}^\chi(r,z) = \hat{D}_z^- \hat{D}_z^+ \varphi_{2}^\chi(r,z)
\end{aligned}    
\right.,
\end{equation}
where we define the longitudinal operators 
\begin{equation}
    \hat{D}_z^{\pm} = i \left[\chi \lambda \pm (i\partial_z+\chi b)\right],
\end{equation}
possessing the properties listed below.
 \begin{itemize}
    \item $\hat{D}_z^+ \hat{D}_z^- = \hat{D}_z^- \hat{D}_z^+  = -\left[\partial_z^2 - 2i\chi b \partial_z +\lambda^2 -b^2\right]$.
    \item $[\hat{D}_z^+,\hat{D}_z^-]= 0$.
    \item $[\hat{D}_r^{\pm},\hat{D}_z^{\pm}]= 0$.
\end{itemize}

As a result of the last property, i.e. commutation between $\hat{D}_r$ and $\hat{D}_z$ operators, arising as their differential operators act on different variables, decoupling the two linear equations in (\ref{eq:decoupled_finite}) results in two second-order differential equations that can be separated over its variables $r$ and $z$. With this in mind, we can propose spinor components with separated variables, that is
\begin{equation}
    \varphi^{\chi} = \begin{pmatrix}
        \varphi_1^{\chi}\\
        \varphi_2^{\chi}
    \end{pmatrix}  =  \begin{pmatrix}
        R^{\chi}_{1}(r)Z_{1}^{\chi}(z) \\
        R^{\chi}_{2}(r)Z_{2}^{\chi}(z)
    \end{pmatrix}.
\end{equation}
Using this form of the spinor, we obtain two equations of separate variables, one for the $r$ and other for $z$, for each chirality, as follows:
\begin{enumerate}
    \item Equation in $r$:
    \begin{equation}
\left[r^2 \partial_r^2  + r\partial_r  +  r^2(\lambda^2 - \kappa^2) -(j' \mp 1/2)^2 \right] R^{\chi}_{1,2} = 0.
\end{equation}
This is again a Bessel equation, so that solutions are given by
\begin{equation}
R^{\chi}_{1}(r) = A_{1}^\chi J_{\nu_-}(\alpha r) + B_{1}^\chi Y_{\nu_-}(\alpha r),
\end{equation}
\begin{equation}
R^{\chi}_2(r) = A_{2}^\chi J_{\nu_+}(\alpha r) + B_{2}^\chi Y_{\nu_+}(\alpha r),
\end{equation}
where $\alpha = \sqrt{\lambda^2 - \kappa^2}$ and where the orders, $\nu_\pm=j'\mp1/2$, are the same defined in the infinite case. $\kappa$ is the constant of separation, and is obtained from the BCs in $z$-edges. Note that now the solutions depend on the direction of propagation of the plane waves in $z$-direction, additionally to the chirality index $\chi$.

    \item Equation in $z$:
    \begin{equation}\label{eq:finite_z}
\left[\partial_z^2  - 2i\chi b \partial_z  - (b^2 - \kappa^2)\right] Z^{\chi}_{1,2} = 0.
\end{equation}
\end{enumerate}

A convenient way to remove the first-derivative term is to factor out a plane wave phase associated with the node separation $b$, that is
\begin{equation}
    Z_{1,2}^\chi(z)= e^{i\chi b z}\tilde{Z}_{1,2}^\chi(z).
\end{equation}
This phase acts with different sign for each chiral node, reflecting contra-propagation between Weyl cones states. After this, the differential equation in
(\ref{eq:finite_z}) takes the form of a Helmholtz equation
\begin{equation}
    \left[\partial_z^2   + \kappa^2\right] \tilde{Z}^{\chi}_{1,2} = 0,
\end{equation}
which has as solutions linear combinations of plane waves propagating in positive and negative $z$-directions, that is
\begin{equation}
\tilde{Z}^{\chi}_{1,2}(z) = \tilde{Z}^{\chi}_\xi(z) = \sum_{\xi=\pm} A_\xi^\chi e^{i\xi\kappa z},
\end{equation}
so that the $Z(z)$ functions have the general form
\begin{equation}
Z^{\chi}_{1,2}(z) = e^{i\chi bz} \sum_{\xi=\pm} A_\xi^\chi e^{i\xi\kappa z}.
\end{equation}
Note that the radial solutions, and consequently the total solutions, will inherit the $\xi$ index. We require that the first-order Dirac equation Eq. (\ref{eq:Dirac}) be satisfied, 
\begin{equation}
    \left[\partial_r +\frac{1}{r}\left(j'+\frac{1}{2}\right)\right] \varphi_2^\chi(r,z) = i(\chi\lambda-\xi\kappa) \varphi_1^\chi(r,z),
\end{equation}
\begin{equation}
    \left[\partial_r -\frac{1}{r}\left(j'-\frac{1}{2}\right)\right] \varphi_1^\chi(r,z) = i(\chi\lambda+\xi\kappa) \varphi_2^\chi(r,z),
\end{equation}
so that, using again the standard Bessel identities in Eqs. (\ref{eq:Bessel_Id1}) and (\ref{eq:Bessel_Id2}), and as $J_{\nu_\pm}$ and $Y_{\nu_\pm}$ are linearly independent, we find the coefficients 
\begin{equation}
    A_{2,\xi}^\chi = i \frac{\chi\lambda - \xi \kappa}{\alpha} A_{1,\xi}^\chi,
\end{equation}
and
\begin{equation}
    B_{2,\xi}^\chi = i \frac{\chi\lambda - \xi \kappa}{\alpha} B_{1,\xi}^\chi.
\end{equation}
Putting it all together, the complete stationary eigenmode is written explicitly as
\begin{equation}
    \Psi\chi(r,\theta,z)=e^{i(j-\frac{1}{2})\theta} \begin{pmatrix}
        \varphi_1^\chi(r,z)  \\ \varphi_2^\chi(r,z) e^{i\theta}
    \end{pmatrix},
\end{equation}
where
\begin{equation}
    \varphi_1^\chi(r,z) = e^{i\chi b z} \sum_{\xi=\pm} e^{i\xi \kappa z}\left[A_{1,\xi}^\chi J_{\nu_-}(\alpha r) + B_{1,\xi}^\chi Y_{\nu_-}(\alpha r)\right],
\end{equation}
\begin{equation}
    \varphi_2^\chi(r,z) = e^{i\chi b z} \sum_{\xi=\pm} e^{i\xi \kappa z} i \beta_{\chi,\xi}\left[A_{1,\xi}^\chi J_{\nu_+}(\alpha r) + B_{1,\xi}^\chi Y_{\nu_+}(\alpha r)\right],
\end{equation}
and
\begin{equation}
    \beta_{\xi,\chi}=\frac{\chi\lambda - \xi \kappa}{\alpha}.
\end{equation}

At this point, spinorial solutions are completely general. In order to find the value of coefficients $A_{1,\xi}^\chi$, $B_{1,\xi}^\chi$, we have to impose the corresponding BCs over the solutions. As the two chiral cones are separated by a distance $2b$ in $z$-direction, we assume no inter-valley scattering at the radial edges. In other words, a particle in a cone with defined chirality (say $\chi=-$) needs a big momentum transfer from the radial walls, $\Delta k \sim |K_- - K_+|$, to reach the other cone with opposite chirality ($\chi=+$ in this case). We are considering here that the radial walls of the cylinder do not provide enough momentum in $z$-direction to present this phenomenon. On the other hand, inter-valley scattering is obviously present at both caps of the cylinder $z=0$ and $z=L$. In terms of BCs, the last discussion translates as assuming IMBCs at the radial walls (similar to the infinite cylinder case which do not mix chiralities), while employing MIT bag BCs \cite{chodos1974new,berry1987neutrino} at longitudinal caps which mix chiralities (inter-valley scattering) \cite{baireuther2015scattering,akhmerov2008boundary}.

Let us start with the IMBCs at the cylinder radial edges at $r=r_1$ and $r=r_2$. This BCs are applied analogously to the infinite case, that is, using Eq. (\ref{eq:r_conditions}). For the inner radius $r=r_1$:
\begin{equation}
\varphi_1^\chi = -i\chi e^{-i\theta} \varphi_2^\chi \bigg|_{r=r_1},
\end{equation}
which yields
\begin{align}
&\sum_{\xi=\pm} e^{i\xi q z} \left[ A_{1,\xi}^\chi J_{\nu_-}(\alpha r_1) + B_{1,\xi}^\chi Y_{\nu_-}(\alpha r_1) \right]\nonumber \\ &= -i\chi \sum_{\xi=\pm} e^{i \xi q z} \left[ A_{1,\xi}^\chi J_{\nu_+}(\alpha r_1) + B_{1,\xi}^\chi Y_{\nu_+}(\alpha r_1) \right] i \beta_{\xi,\chi}.
\end{align}
As the functions $e^{i\xi \kappa z}$ are linearly independent, we have:
\begin{align}\label{eq:radial_cond_1}
	& A_{1,\xi}^\chi [J_{\nu_-}(\alpha r_1) -\chi\beta_{\xi,\chi} J_{\nu_+}(\alpha r_1)] \nonumber \\ &+ B_{1,\xi}^\chi [Y_{\nu_-}(\alpha r_1) - \chi\beta_{\xi,\chi} Y_{\nu_+}(\alpha r_1)] = 0 .
\end{align}

Proceeding in a similar way for the outer radius $r=R_2$: 
\begin{equation}
    \varphi_1^\chi = i\chi e^{-i\theta} \varphi_2^\chi \big|_{r=r_2}
\end{equation}
reads
\begin{align}\label{eq:radial_cond_2}
 & A_{1,\xi}^\chi [J_{\nu_-}(\alpha r_2) + \chi\beta_{\xi,\chi} J_{\nu_+}(\alpha r_2)] \nonumber \\ &+ B_{1,\xi}^\chi [Y_{\nu_-}(\alpha r_2) +\chi \beta_{\xi,\chi} Y_{\nu_+}(\alpha r_2)]  = 0.
\end{align}
This can be written in matrix form similar to the infinite case, with the condition $\det(\mathbb{M})=0$ determining a transcendental equation for the energy levels, which will now depend on $\chi, \xi$ (and a transversal quantum number, $l$, resulting from the quantization of the transversal momentum, as we will see next). Also, from Eqs. (\ref{eq:radial_cond_1}) or (\ref{eq:radial_cond_2}) constants $B_{1,\xi}^\chi$ can be written in terms of $A_{1,\xi}^\chi$.

\begin{figure*}
    \centering
    \includegraphics[width=0.95\textwidth]{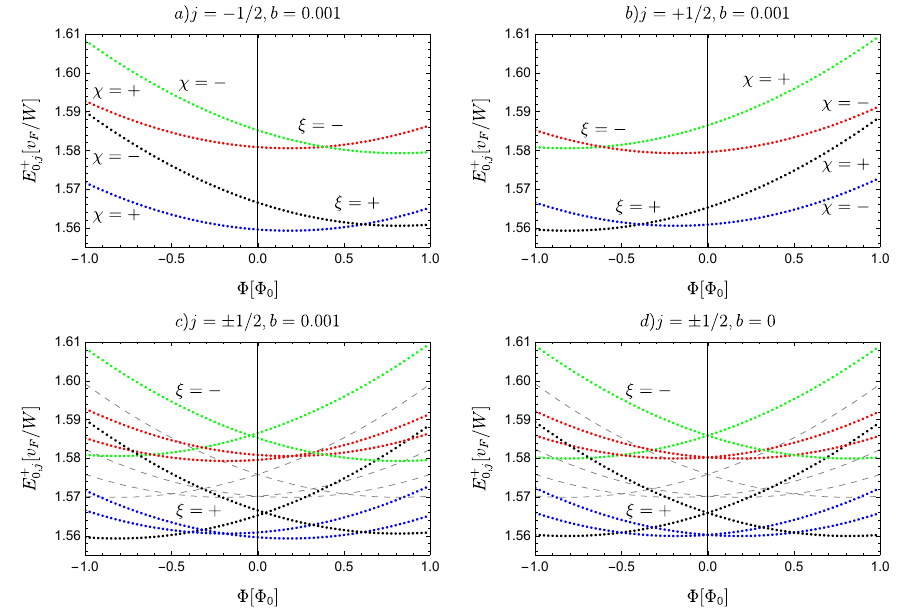}
    \caption{Energy eigenvalues as a function of the magnetic flux, $\Phi$, for the angular momentum, $j=\pm\frac{1}{2}$, for the finite cylinder. $a$) and $b$) are computed for $j=-\frac{1}{2}$ and $j=\frac{1}{2}$, respectively, for the case $b=0.001$, $l=0$; $c$) shows both $j=\pm\frac{1}{2}$ for the same case. $d$) Shows both $j=\pm\frac{1}{2}$ for the case $l=0$, $b=0.001$. In both cases, energies for states with both chiralities, $\chi=\pm$, and both propagation directions, $\xi=\pm$, are presented. In all the figures in the paper we have used the values $r_1=5.5$, $r_2=6.5$, and for the finite case we used additionally the value $L=100$.}
    \label{fig:finite_energy_kn0_bn0_bo}
\end{figure*}

Let us now apply chiral BCs over the longitudinal caps at $z=0$ and $z=L$ \cite{berry1987neutrino,chodos1974new},
\begin{equation}
    \Pi_+ \Psi(r,\theta,z)\bigg|_{z=0,L}=0,
\end{equation}
where the chiral projector $\Pi_+$ is defined as
\begin{equation}
    \Pi_+= \frac{1}{2} \left(\mathbb {I}_{4\times 4} +i \gamma^n\right),
\end{equation}
where $\gamma^n = n_k\gamma^k$, with $n_k$ the normal direction of the corresponding boundary and $\gamma^k$ the Dirac matrices in chiral (Weyl) representation.
At $z=0$ the value $n_z=-1$ is taken, resulting in the condition
\begin{equation}
    \frac{1}{2} \left(\mathbb {I}_{4\times 4} -i \gamma^z\right)\Psi(r,\theta,0)=0.
\end{equation}
From the last expression, two conditions arise
\begin{equation}
    \varphi_1^-(r,\theta,0) = -i\varphi_1^+(r,\theta,0), \quad \varphi_2^-(r,\theta,0) = i\varphi_2^+(r,\theta,0).
\end{equation}
Explicitly,
\begin{subequations}
\begin{align}\label{eq:BCs_z1}
     &A_{1+}^- J_{\nu_-} + B_{1+}^- Y_{\nu_-} + A_{1-}^- J_{\nu_-} + B_{1-}^- Y_{\nu_-} = \nonumber \\ &-i \left[ A_{1+}^+ J_{\nu_-} + B_{1+}^+ Y_{\nu_-} + A_{1-}^+ J_{\nu_-} + B_{1-}^+ Y_{\nu_-} \right],
\end{align}
and
\begin{align}\label{eq:BCs_z2}
&i \beta_{-+} (A_{1+}^- J_{\nu_+} + B_{1+}^- Y_{\nu_+}) + i \beta_{--} (A_{1-}^- J_{\nu_+} + B_{1-}^- Y_{\nu_+}) = \nonumber  \\ &i \left[ i \beta_{++} (A_{1+}^+ J_{\nu_+} + B_{1+}^+ Y_{\nu_+}) + i \beta_{+-} (A_{1-}^+ J_{\nu_+} + B_{1-}^+ Y_{\nu_+}) \right].
\end{align}
\end{subequations}
From Eq. (\ref{eq:BCs_z1}) we arrive to the conditions over the constants
\begin{equation}
A_{1,+}^- = -iA_{1,-}^+, \qquad  A_{1,-}^- = -iA_{1,+}^+. 
\end{equation}
From Eq. (\ref{eq:BCs_z2}) we obtain
\begin{equation}
    A_{1,-}^+ = -A_{1,+}^+ \frac{\beta_{++}+\beta_{--}}{\beta_{+-}+\beta_{-+}}.
\end{equation}
Similar relations are found for $B_{1,\xi}^\chi$ constants. At this stage we have completely defined all the parameters $A_{1,\xi}^\chi$, $A_{2,\xi}^\chi$, $B_{1,\xi}^\chi$, $B_{2,\xi}^\chi$, in terms of a unique constant, $A_{1,+}^+$, that plays the role of the normalization constant.

At $z=L$, $n_z=+1$, the BCs result in the two conditions 
\begin{equation}\label{eq:cond_L}
    \varphi_1^-(r,\theta,L) = i\varphi_1^+(r,\theta,L), \quad \varphi_2^-(r,\theta,L) = -i\varphi_2^+(r,\theta,L).
\end{equation}

From these conditions,  a quantization over the transversal momentum is obtained
\begin{equation}
     \kappa_{l,\chi}\equiv k_{z,l,\chi}= \left(\frac{2l+1}{2}\right) \frac{\pi}{L}+\chi b, \quad l\in\ \mathbb{Z},
\end{equation}
therefore, $\kappa_{l,\chi}-\chi b$ is always an odd multiple of $\pi/2L$. It is important to note out that the quantized form of $\kappa$, that depends on the chirality, enters in the transcendental equation of the energy spectrum of the system, causing an energy shift for each chiral node, similar to the case of the infinite cylinder, but for discrete values of $\kappa$. This asymmetry is the signature of a system where time reversal or parity is broken, which is the essence of a WSM with nodes separated by $2b$.

\subsection{Energy spectrum}\label{sec:energy_finite}

As already mentioned, the condition $\det(\mathbb{M}) = 0$ determines a transcendental
equation for the energy levels, which  now depends on the chirality index
$\chi$, the propagation direction $\xi$, and the transversal quantum number $l$. In fact, the parameter $\beta_{\xi,\chi}=\beta_{\xi,\chi,l}$, that is, also depends on $l$, while $\alpha=\alpha_{\chi,l}$, that is, depends on $\chi,l$, so that we have an energy spectrum which depends on $\xi,\chi,l$. Figure \ref{fig:finite_energy_kn0_bn0_bo} shows the energy spectrum as a function of the magnetic flux, $\Phi$, for the angular momentum $j=\pm\frac{1}{2}$.

Figures \ref{fig:finite_energy_kn0_bn0_bo}$a$) and \ref{fig:finite_energy_kn0_bn0_bo}$b$) show the spectrum for $j=-\frac{1}{2}$ and $j=\frac{1}{2}$, respectively, for the case $b=0.001$, $l=0$. It can be seen an unfolding of the energy spectrum induced by the transversal propagation direction parameter $\xi$, that is, particles with positive transversal direction shift their energy to lower values in the spectrum, while particles with negative transversal direction shift toward greater energies, defining two branches, one for each propagation direction. Also, similar to the infinite case, inside each $\xi$-branch a shift in energy is induced by the separation between nodes, $b\neq 0$. Considering these two effects inside the finite cylinder, we end up with a total energy spectrum (considering all $\xi$'s and all $\chi$'s) as the shown in Fig. \ref{fig:finite_energy_kn0_bn0_bo}$c$). We observe an asymmetry in the total energy
spectrum resulting from these two interactions, presenting a TR symmetry breaking in the material
even when no net magnetic flux is present, similar to the infinite case. It should also be noted that the spectrum for the $\xi=-$ branch is exactly the inverted spectrum for the $\xi=+$ branch with respect to the $\Phi=0$ axis. In contrast, from Fig. \ref{fig:finite_energy_kn0_bn0_bo}$d$) can be seen that when there is no node separation, $b=0$, the transversal propagation still induces two energy branches for $\xi=\pm$, however, no asymmetry between energy levels at each branch occur, and consequently no TR symmetry breaking occurs for $\Phi=0$. As a
consequence, chiral degeneracy for $\Phi=0$ is lifted in the finite case for each propagation direction only in the presence of a finite separation of the chiral nodes.

Furthermore, the introduction of longitudinal confinement in the finite cylinder reveals a subtle interplay between boundary effects and valley physics. By imposing MIT bag BCs at the cylinder caps ($z=0, L$), we explicitly permit inter-valley scattering, a physical mechanism strictly absent in the infinite case under purely radial confinement. These longitudinal boundaries act as hard topological walls that mix the chiral states, coupling the incoming wave of one chirality with the reflected wave of the opposite chirality. This boundary-induced inter-valley scattering is the fundamental physical mechanism responsible for the $\xi$-dependent unfolding of the energy branches. As a result, the propagation direction $\xi$ becomes strongly hybridized with the chiral properties of the bulk, leading to the highly asymmetric and complex energy spectrum observed when both the internal axial field ($b \neq 0$) and the external AB flux ($\Phi$) are present.


\subsection{Persistent current inside the cylinder}

In this case, as we have just seen in Sec. \ref{sec:energy_finite}, even when the energy spectrum decouples in two branches depending on the transversal propagation direction $\xi$, inside each branch the spectrum is similar to that of the infinite case (Fig. \ref{fig:infinite_persistent}), in which for $b=0$ a vanishing persistent current is found, but a $b\neq 0$ generates a finite persistent current
at $\Phi=0$, that is, the system presents a finite persistent
current even without magnetic flux, exclusively induced by
the separation of the chiral nodes. In Addition, if we compute the persistent current for each branch it could be observed that the curves present the same structure with the changes $\Phi \longrightarrow -\Phi$, $j_{\xi}\longrightarrow -j_{-\xi}$. This has great relevance in the conductance channels of the system, as will be seen next.


\begin{figure}
    \centering
    \includegraphics[width=0.95\columnwidth]{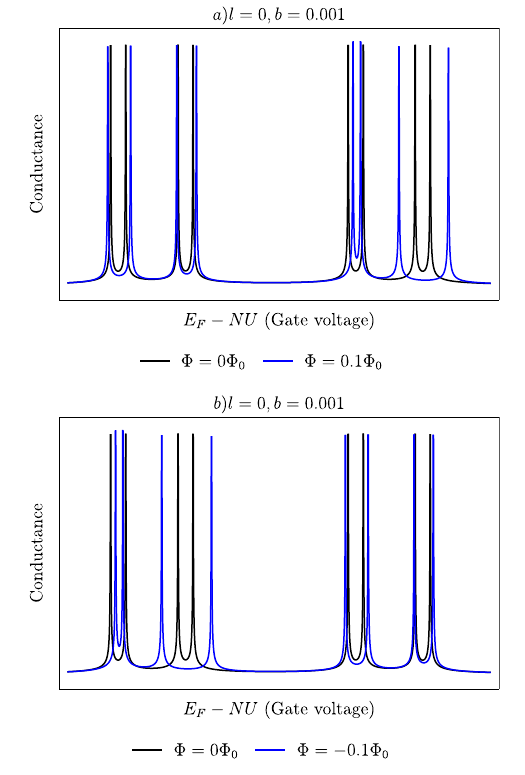}
    \caption{Conductance resonances (conductance channels) through the finite thick-walled cylinder coupled to leads as a function of the Fermi energy (gate voltage), for for $l = 0$, $b = 0.001$. $a$) Shows the cases $\Phi=0$ (black lines) and $\Phi=0.1$ (blue lines). $b$) Shows the cases $\Phi=0$ (black lines) and $\Phi=-0.1$ (blue lines).}
    \label{fig:finite_cond}
\end{figure}


\subsection{Conductance}

We computed the conductance resonances (channels) for the finite cylinder system in Fig. \ref{fig:finite_cond} for $l=0$, $b=0.001$, for magnetic flux $\Phi=0$ (black lines), $\Phi=0.1$ (Fig. \ref{fig:finite_cond}$a$), blue lines), and $\Phi=-0.1$ (Fig. \ref{fig:finite_cond}$b$), blue lines). First, comparing to the infinite case in which the chiral degeneracy is broken, resulting in the unfolding of the fourfold degenerate conductance channels into two twofold degenerate channels, in this case each twofold degenerate channel is additionally unfolded into two no degenerated conductance channels as result of the transversal propagation non-equivalence. Therefore, each fourfold degenerated channel in the graphene case in now unfolded into four no degenerated channels for each $\chi=\pm$, $\xi=\pm$. Secondly, as comparing Figs. \ref{fig:finite_cond}$a$) and \ref{fig:finite_cond}$b$), it can be noted that the channels for $\Phi=-0.1$ are basically the same as for $\Phi=-0.1$, but where the conductances between $\xi$ branches are exchanged for those with opposite $\xi$. In other words, the first eight channels are exchanged for the second group of eight channels when inverting the magnetic flux in the material.

\begin{figure*}
    \centering
    \includegraphics[width=0.99\textwidth]{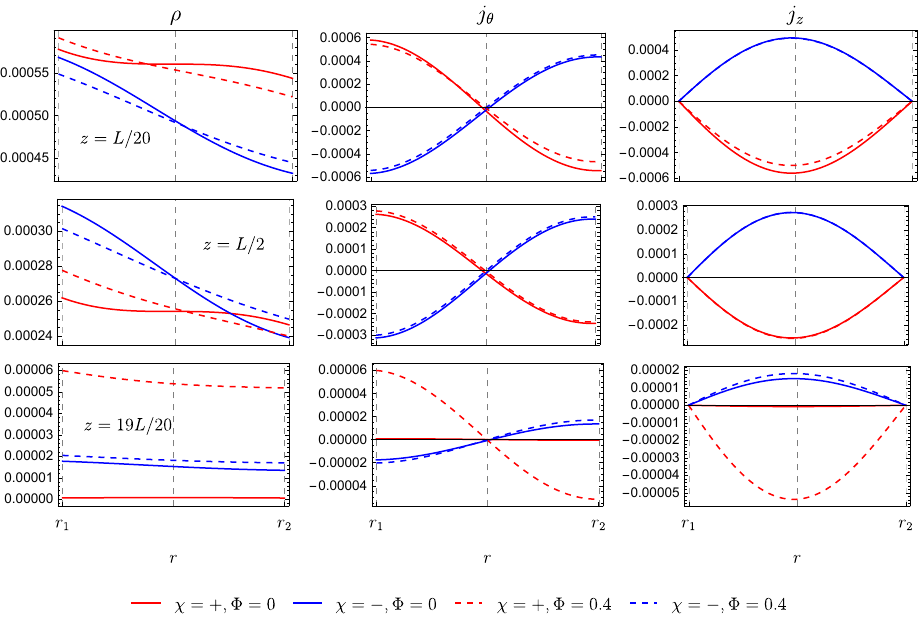}
    \caption{Probability density and currents as a function of $r$ for particles with defined chirality $\chi=+$ (red lines) and $\chi=-$ (blue lines) for the
finite cylinder problem. Solid lines refer to the case with no magnetic flux ($\Phi=0$) in the cylinder, while dashed lines
correspond to finite magnetic flux ($\Phi=0.4$). First to third rows depict the cases when we fix $z=L/20$, $L/2$ and $19L/20$, respectively. As there is no angular dependence we used $\theta=0$.}
    \label{fig:finite_currents}
\end{figure*}

\subsection{Density probabilities and currents}

Unlike the infinite case where the only dependence of the probability density and currents was in the radial coordinate, in the finite case the transversal $z$ coordinate no longer presents a constant behavior. Figure \ref{fig:finite_currents} shows the probability density, as well as the current densities $j_\theta$, $j_z$ as a function of the radial coordinate, for both chiralities (red lines for $\chi=+$ and blue lines
for $\chi=-$, respectively) and for zero and finite magnetic
flux (solid lines for $\Phi=0$ and dashed lines for $\Phi=0.4$,
respectively), for $z=l/20,L/2,19L/20$ (first to third row, respectively). 

Near the inferior edge, at $z=L/20$, the probability density shows localization of chiral particles around the internal cylinder radius, $r_1$. Also, particles with $\chi=+$ present more concentration for all $r$, that is, the system presents chiral imbalance for this value of $z$ (not for the whole cylinder). The effect of the magnetic flux, $\Phi$, is to ``linearize'' the probability density curves, but keeping the localization at the radial borders. On the other hand, the angular and transversal currents are opposite for each chirality presenting a behavior very similar to the infinite case, with the difference that the magnetic flux acts with greater weight for particles with positive chirality.

At the center of the cylinder, $z=L/2$, the probability density shows a localization localization of chiral particles similar to $z=L/20$, but with major concentration of $\chi=-$ particles, where a crossing point exists between chiral curves, that is, a chiral balance is present there, but in general a chiral imbalance is present. Just as the $z=L/20$ case, the effect of the magnetic flux, $\Phi$, is to ``linearize'' the probability density curves. In this case, even though we are at the center of a very long cylinder ($L\gg W$), the probability density is different from the infinite case due to the range considered [$0,L$], which induces phases over the spinor. Results similar to the infinite case can be consistently recovered considering a symmetric range (e.g. [$-L/2,L/2$]). The angular and transversal currents are again very similar to the infinite case, with the magnetic flux affecting minimally  both chiral particles equally. 

The most dramatic behavior appears to be near the top edge, $z=19L/20$. Here, the probability density presents a weak localization around the internal radius, decreasing with a quasi-linear behavior as moving to the external radius. For $\Phi=0$ we have chiral imbalance with a greater concentration of $\chi=-$ particles (in fact, the probability density for $\chi=+$ particles is practically zero). However, for a finite value of $\Phi$, $\chi=+$ probability density is mostly affected, inverting the chiral imbalance. Something similar occurs with the currents $j_\theta$ and $j_z$. They present a finite $\chi=-$ current and a practically vanishing $\chi=+$ current for $\Phi=0$. However,  $\chi=+$ currents are greatly influenced by the magnetic flux (not so the $\chi=-$ currents), scaling to values greater than the opposite chiral current. 

In addition to the last analysis, it is important noting that the scales in the plots of Fig. \ref{fig:finite_currents} decreases as we go higher in the cylinder, reducing an order of magnitude each quantity near the top compared to near the bottom of the cylinder. Also, even though the radial current $j_r$ presents some non constant behavior, we do not show it as its magnitude is very small ($\sim 10^{-7}$). In this finite case, the probability density and current densities
remain constant for all  and $\theta \in [0, 2\pi]$ for $r, z$
fixed, so that we choose $\theta=0$ in all our calculations.

\begin{figure*}
    \centering
    \includegraphics[width=\textwidth]{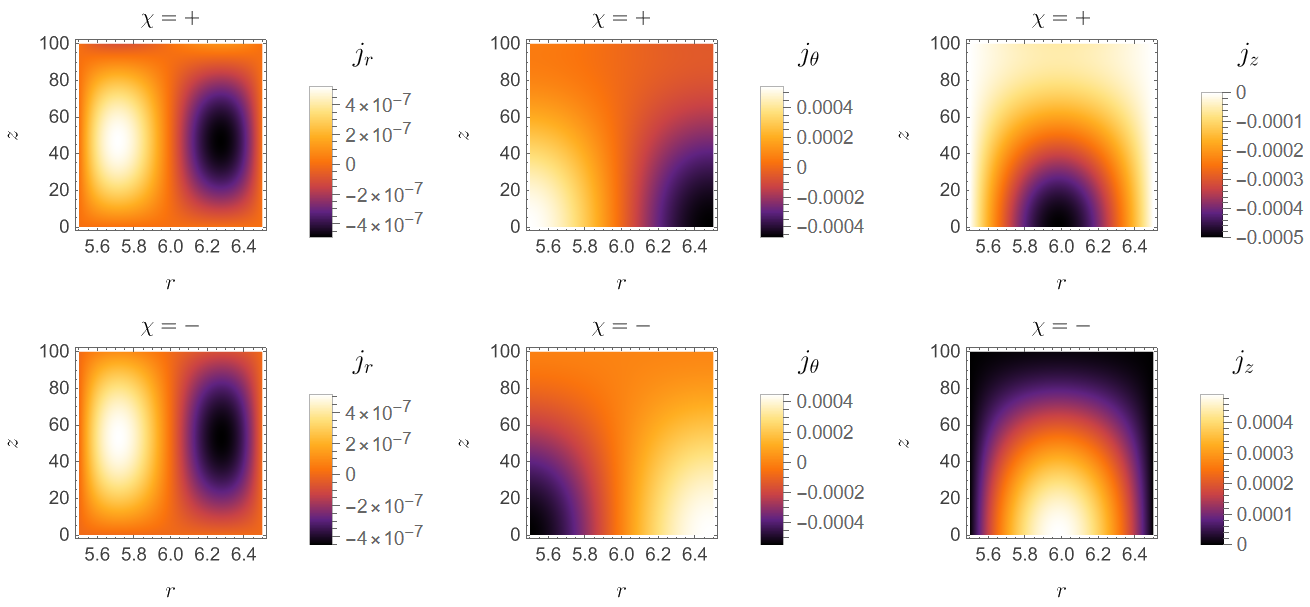}
    \caption{Density plots of the chiral density currents for the finite cylinder.}
    \label{fig:finite_currents_density}
\end{figure*}

\begin{figure*}
    \centering
    \includegraphics[width=0.95\textwidth]{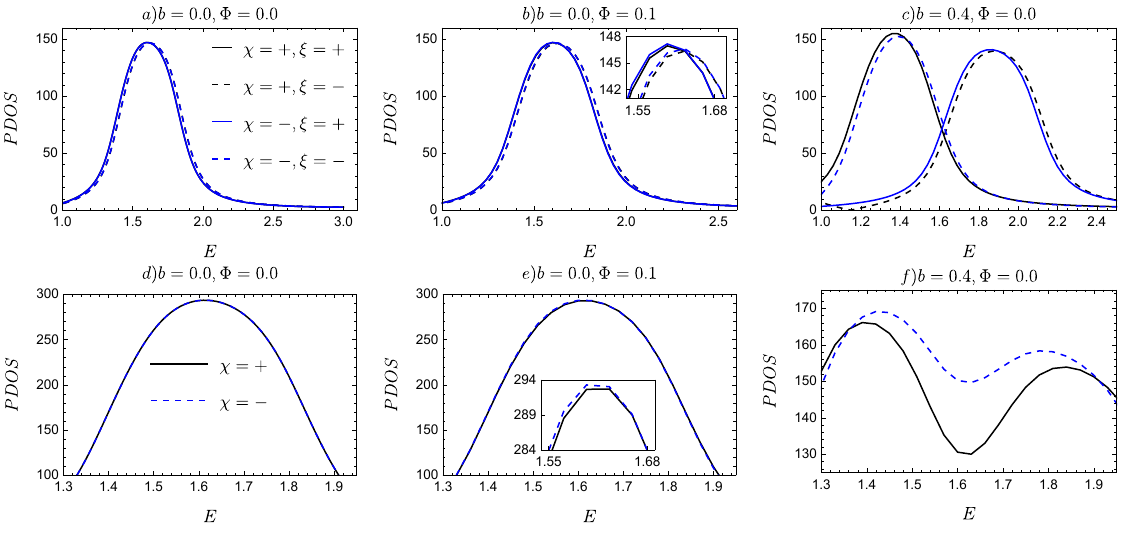}
    \caption{Partial density of states (PDOS) for the finite cylinder. $a$)-$c$) include computed PDOS for each value of $\chi=\pm$ (black and blue lines, respectively) and $\xi=\pm$ (solid and dotted lines, respectively). $d$)-$f$) correspond to PDOS for each chirality (black solid lines and dotted blue lines for $\chi=\pm$, respectively), summing up the propagation direction for each case.}
    \label{fig:finite_DOS}
\end{figure*}

Figure \ref{fig:finite_currents_density} shows a density plot of all the current densities, $j_r$, $j_\theta$ and $j_z$, as a function of the $r$, $z$ coordinates for $\Phi=0$. We observe  counter-propagating angular currents for different chiralities, similar to the AB effect, induced by the separation of the chiral nodes $b$. We also see counter-propagating transversal currents, while radial currents propagating with the same direction, but with very small magnitudes, as mentioned in the last paragraph. Finally, these figures indicate a helical motion of chiral particles with opposite directions for each chirality, and depending on the value $r$, that is, the motion inverts near the center of the thick- walled cylinder, $(r_1+r_2)/2$.

\subsection{Density of states}

Figure \ref{fig:finite_DOS} shows the calculated partial density of states (PDOS) for the finite cylinder. Figs. \ref{fig:finite_DOS}$a$)-$c$) include computed PDOS for each value of $\chi=\pm$ (black and blue lines, respectively) and $\xi=\pm$ (solid and dotted lines, respectively). Figs. \ref{fig:finite_DOS}$d$)-$f$) correspond to PDOS for each chirality (black solid lines and dotted blue lines for $\chi=\pm$, respectively), summing up the propagation direction for each case. When we have no separation of the chiral nodes and no magnetic flux, $b=0$, $\Phi=0$ [Figs. \ref{fig:finite_DOS}$a$),$d$)], we note a degeneration in the chiral index $\chi$ of the PDOS for each propagation direction $\xi$ [Fig. \ref{fig:finite_DOS}$a$)], resulting in degenerated chiral PDOS when summing up across the $\xi$ index [Fig. \ref{fig:finite_DOS}$d$)]. When a finite magnetic flux is added [Figs. \ref{fig:finite_DOS}$b$),$e$)], the chiral degeneration is lifted, that is, TR symmetry breaks. The shifting of the PDOS for each chirality are very small as sown in the insets. Finally, when considering a finite node separation, $b\neq 0$ [Figs. \ref{fig:finite_DOS}$c$),$f$)], each chiral PDOS widely unfolds for each propagation direction. This is consistent with the energy separation for each propagation direction branch calculated in Sec. \ref{sec:energy_finite}, where the energies with $\xi=+$ shift to lower energies, while the energy values with $\xi=-$ shift to higher energy values [Fig. \ref{fig:finite_DOS}$c$)]. This results in unfolded PDOS for each chirality, i.e. TR symmetry breaks. In conclusion, similar to the infinite case, the system presents two mechanisms to break TR symmetry, one by means of a node separation of the chiral nodes, and other by the effects of an external magnetic flux (AB effect), being the propagation direction of the chiral particles of notable importance in the finite cylinder case.

\section{Conclusions}
\label{sec:conclusions}
In this work, we have provided a comprehensive analytical study of the electronic structure, transport properties, and persistent currents in a thick-walled Weyl semimetal cylinder pierced by an AB flux. By treating both infinite and finite-length geometries, we have elucidated the profound interplay between the intrinsic topological properties of the WSM (specifically, the spatial separation of the chiral nodes in momentum space) and the external magnetic flux.  For the infinite cylinder, governed by radial infinite-mass boundary conditions, we found that the node separation parameter $b$ effectively acts as an internal chiral gauge field. This intrinsic field is sufficient to break time-reversal symmetry globally, resulting in a lifting of the chiral degeneracy even at $\Phi = 0$. Consequently, the system exhibits spontaneous, dissipationless persistent currents and an unfolding of the fourfold degenerate conductance channels into twofold degenerate ones, establishing a purely topological signature independent of the external AB flux.  The transition to a finite-length cylinder introduces rich mesoscopic physics driven by longitudinal confinement. By implementing MIT bag boundary conditions at the cylinder caps, which allow for inter-valley scattering, we demonstrated a quantization of the longitudinal momentum that strongly depends on the propagation direction $\xi$. This geometric confinement induces an additional splitting in the energy spectrum and the PDOS, fully unfolding the conductance resonances into non-degenerate channels. Furthermore, our analysis of the local probability densities and spatial currents reveals a complex helical motion of chiral particles. We found that the finite boundaries induce strong spatial chiral imbalances, particularly near the cylinder ends, where the magnetic flux can invert the dominant chirality of the localized states.  From an experimental standpoint, our results suggest that mesoscopic thick-walled WSM cylinders (or tubular nanowires) offer a highly tunable platform for observing anomalous topological transport. The predicted spontaneous persistent currents at zero flux, alongside the high-resolution splitting of conductance channels driven by finite-size effects, provide clear, measurable macroscopic manifestations of the underlying chiral anomaly and Weyl physics in confined geometries.

\section*{Acknowledgments}
J.C.P.-P was supported by the SECIHTI under the program ``Estancias Posdoctorales por México'' with CVU
number 671687. J.A.C. gratefully acknowledges the support of SECIHTI through the program \textit{Becas Nacionales para estudios de Posgrado}, under grant number 4018746. D.A.B. was supported by the DGAPA-UNAM Posdoctoral Program. A.M.-R. acknowledges financial support by UNAM-PAPIIT project No. IG100224, UNAM-PAPIME project No. PE109226, by SECIHTI project No. CBF-2025-I-1862 and by the Marcos Moshinsky Foundation.

\appendix

\bibliography{biblio}

\end{document}